\documentclass[palatino,10pt]{article}
\usepackage{moriond,epsfig}

\def\Journal#1#2#3#4{{#1} {\bf #2}, #3 (#4)}

\def\PLB{{\em Phys. Lett.}  B}

\def\be{\begin{equation}}
\def\ee{\end{equation}}
\def\bea{\begin{eqnarray}}
\def\eea{\end{eqnarray}}

\begin{document}
\vspace*{4cm}
%
%
%
\newcommand{\OPAL}{\textsf{OPAL}}
\newcommand{\ALEPH}{\textsf{ALEPH}}
\newcommand{\DELPHI}{\textsf{DELPHI}}
\newcommand{\LEP}{\textsf{LEP}}
\newcommand{\LEPI}{\textsf{LEP I}}
\newcommand{\LEPII}{\textsf{LEP2}}
\newcommand{\ATLAS}{\textsf{ATLAS}}
\newcommand{\LHC}{\textsf{LHC}}
\newcommand{\LHCB}{\textsf{LHCb}}
\newcommand{\CERN}{\textsf{CERN}}
\newcommand{\SM}{\textsf{SM}}
\newcommand{\MSSM}{\textsf{MSSM}}
\def \lsim{\mathrel{\mathpalette\@versim<}}
\def \gsim{\mathrel{\mathpalette\@versim>}}
\def\SUXU{{SU(2)_L \times U(1)_Y}}
\def\hAp{(\mh,\mA)}
\def\sba2{\sin ^2 (\beta - \alpha)}
\def\cba2{\cos ^2 (\beta - \alpha)}
\def\tanB{\tan \beta}
\def\r{\rightarrow}
\def\Ecm{E_{CM}}
\def\sqs{\sqrt{s}}
\def\h{\mathrm h}
\def\A{\mathrm A}
\def\W{\mathrm W^{\pm}}
\def\Z{\mathrm Z}
\def\HH{\mathrm H^0_{\mathrm{SM}}}
\def\H{\mathrm H}
\def\Hpm{\mathrm H^\pm}
\def\hAA{\h\r\A\A}
\def\ZhA{\Z\r\h\A}
\def\ZshZ{\Zs\r\h\Z}
\def\ZshA{\Zs\r\h\A}
\def\ZsHSMZ{\Zs\r\H\Z}
\def\ZsHZ{\Zs\r\H\Z}
\def\WW {\mathrm W^+ \mathrm W^-} 
\def\ZZ {\mathrm Z \mathrm Z} 
\def\ee{\mathrm e^+\mathrm e^-}
\def\mm{\mu^{+}\mu^{-}}
\def\nn{\nu \bar{\nu}}
\def\qq{\mathrm q \bar{\mathrm q}}
\def\pb{ \mathrm{pb} ^{-1}}
\def\Gcs{\mathrm{GeV/c}^2}
\def\Tcs{\mathrm{TeV/c}^2}
\def\Mcs{\mathrm{MeV/c}^2}
\def\Gc{\mathrm{GeV/c}}
\def\G{\mathrm{GeV}}
\def\eehad{\mathrm e^+\mathrm e^-\rightarrow \rm{hadrons}}
 
\def\Zs{\mathrm Z^{*}}
\def\tt{\tau^{+}\tau^{-}}
\def\ttqq{$\tau^+-\tau^--\mathrm q-\overline{\mathrm q}~$}
\def\ll{\ell^{+}\ell^{-}}
\def\ff{\mathrm f \bar{\mathrm f}}
\def\cc{\mathrm c \bar{\mathrm c}}
\def\bb{\mathrm b \bar{\mathrm b}}
\def\mtau{m_{\tau}}
\def\mmu{m_{\mu}}
\def\mb{m_{\mathrm b}}
\def\mH{m_{\H}}
\def\mHhat{\hat{m_{\H}}}
\def\mh{m_{\h}}
\def\mA{m_{\A}}
\def\mHH{m_{\HH}}
\def\mZ{m_{\Z}}
\def\mW{m_{\mathrm W}}
\def\mt{m_{\mathrm t}}
\def\mS{m_{\mathrm S}}
\def\pt{p_{t}}
\def\G{ \rm{GeV} }
\def\nhit{\rm{N}_{hit}}
\def\mht{\rm{T^{min}_{hemi}}}
\def\ahm{\rm{m^{avg}_{hemi}}}
\def\hnn{\mathrm H^0\nu\overline{\nu}}
\def\ttnn{$\tau^+\tau^-\nu \bar{\nu}$}
\def\hmm{$\mathrm H^0\mu^+\mu^-$}
\newcommand {\gevp}        {${\rm GeV/c }$}
\newcommand{\epsnn}{\mbox{$\epsilon_{\nu \bar{\nu}}$(\%)}}
\newcommand{\Nnn}{\mbox{N$_{\nu \bar{\nu}}$}}
\newcommand{\epsee}{\mbox{$\epsilon_{e^{+}e^{-}}$(\%)}}
\newcommand{\Nee}{\mbox{N$_{e^{+}e^{-}}$}}
\newcommand{\epsmm}{\mbox{$\epsilon_{\mu^{+}\mu^{-}}$(\%)}}
\newcommand{\Nmm}{\mbox{N$_{\mu^{+}\mu^{-}}$}}
\newcommand{\Nexp}{\mbox{N$_{exp}$}}

\title{THE SM HIGGS BOSON SEARCH at LEP: COMBINED RESULTS}

\author{ Pedro TEIXEIRA-DIAS }

\address{Department of Physics, Royal Holloway University of London, \\
Egham, SURREY TW20 0EX, England}

\maketitle\abstracts{
During the run in the year 2000, with data collected at collision
energies up to 209 GeV, the LEP experiments have possibly unearthed
the first evidence of a Higgs boson signal at $\mh\approx\,115\,\Gcs$.
The preliminary combined results prepared immediately after the end of
the data-taking, in November 2000, are presented here. Overall, a 2.9
$\sigma$ excess over the background is found, consistent with a
Standard Model Higgs boson signal with $\mh=115.0\,\Gcs$.}

\section{Introduction}

 During the year 2000, the LEP collider was pushed to the edge of its
 performance envelope in order to maximise the Standard Model (SM)
 Higgs discovery potential\,\cite{pj}. In total, the ALEPH, DELPHI, L3
 and OPAL experiments have collected $\approx 870\,\pb$ of
 data, mostly around centre-of-mass energies of $\sqrt{s}\approx
 205$ GeV and $\sqrt{s}\approx 206.7$ GeV. Around 60\% of the data was
 collected at $\sqrt{s} > 206$ GeV.

 At 8h00 a.m., November 2nd, the LEP collider was shut down
 forever. On the 3rd of November the experiments presented their
 results at a special CERN seminar\,\cite{ADLONov3}. The combined
 results were also presented\,\cite{PIKNov3}, including the
 quasi-totality ($\simeq 94$\%) of the data taken in 2000. The results
 presented here correspond to this combination. Having been prepared
 during the final days of data-taking, the results are clearly still
 preliminary. At the time of writing, the ALEPH and L3 collaborations
 are preparing their final publications, with DELPHI and OPAL to
 follow, before the end of 2001.

\section{The combination method}

The method for combining the results of the four experiments is
described in detail elsewhere\,\cite{comb-method} and will only be
succinctly introduced here. The combination relies on the extended
likelihood ratio, between the signal+background hypothesis and
the background-only hypothesis (see, e.g., elsewhere in these
proceedings\,\cite{myproc}): 
\[
Q=\frac{L_{s+b}}{L_b}, ~~~-2\ln Q(\mh)=2s_{tot}-2\sum_{i} \ln(1+(s/b)_i)  
\]
where $s_{tot}$ is the expected number of signal events in a given
channel (search topology, $\sqrt{s}$, and experiment). Each of the
data candidates $i$ contributes a term with a weight $\ln(1+(s/b)_i)$
to $-2\ln Q$. $(s/b)_i$ is the local signal-to-background ratio for
the given candidate, and is determined from the signal and background
p.d.f.s for reconstructed mass as well as additional discriminant
information (e.g., b-tagging, NN output). In order to combine the
various channels one adds up the respective $-2\ln Q$ values.

\section{The combined results}

\begin{figure}[htb]
\epsfig{file=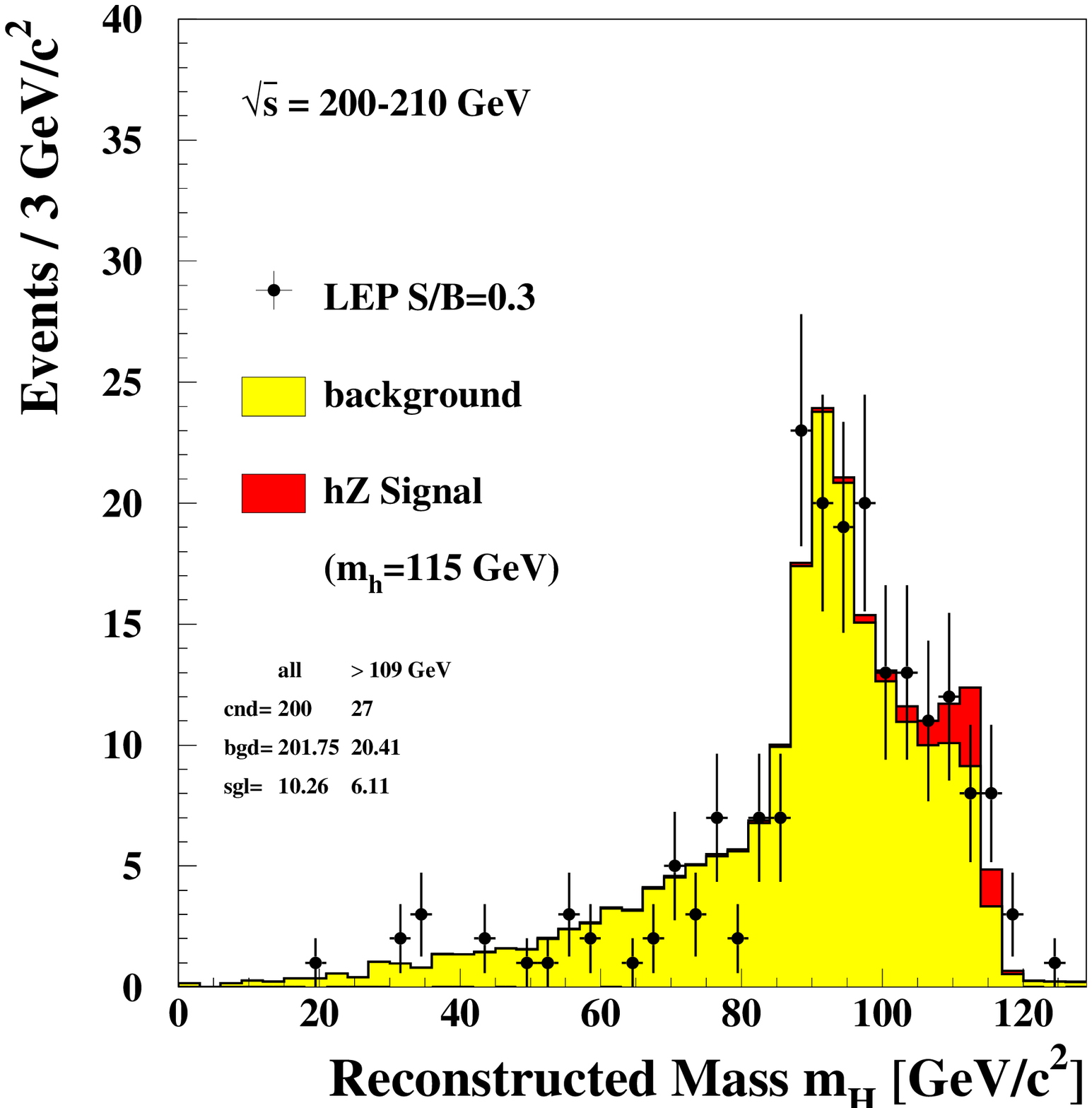,width=7.5cm}\hspace{0.5cm}\epsfig{file=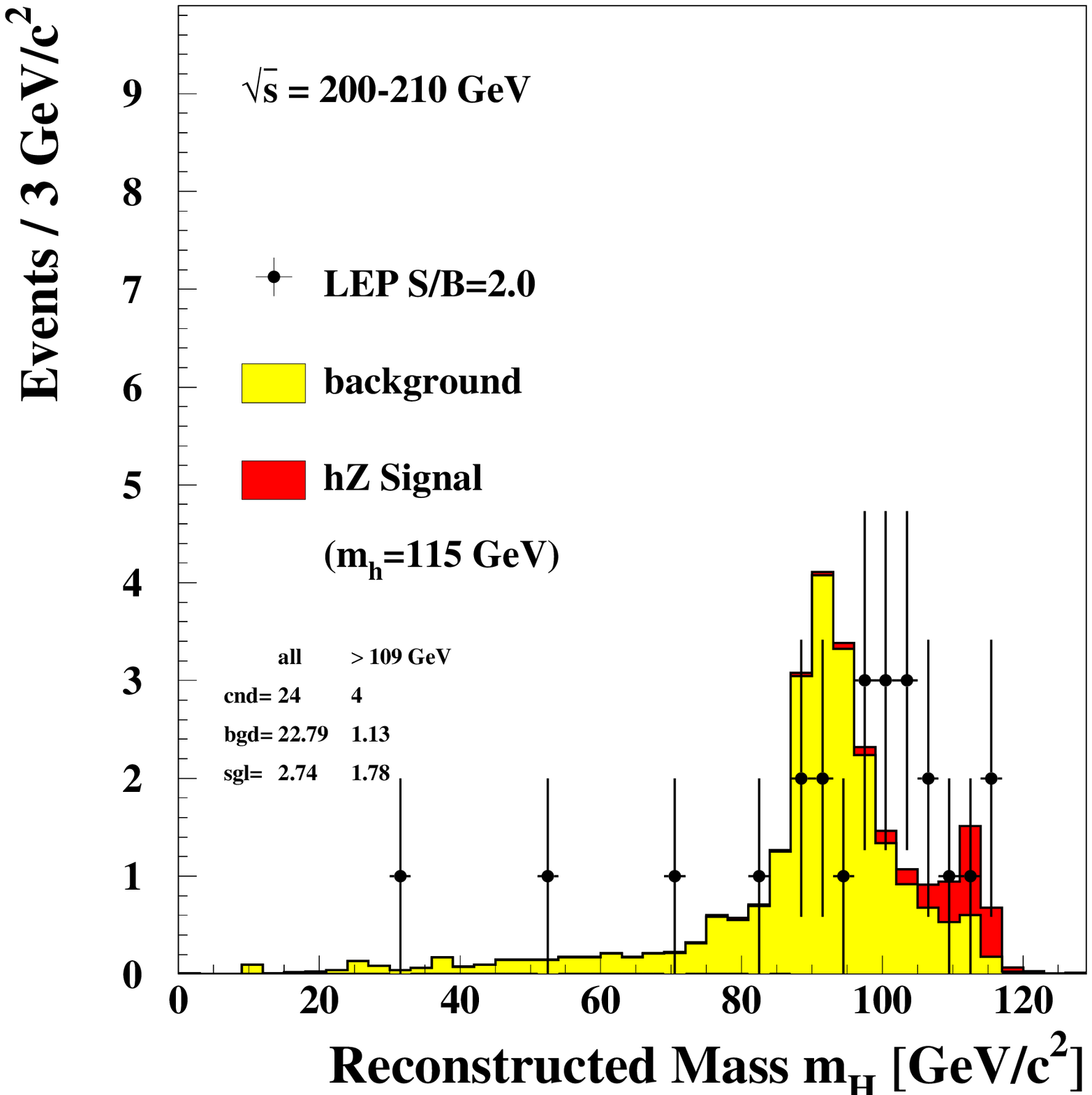,width=7.5cm}\\
\caption{
Reconstructed mass distributions for the candidate events selected by
the four LEP experiments in the data collected in the year 2000. The
distributions are shown at two different selection levels (see text):
(a) loose event selection, low purity for $\mh=115\,\Gcs$; (b) tight
event selection, yielding a high-purity subsample of events for
$\mh=115\,\Gcs$.\label{fig:massplots}}
\end{figure}

 
 The distribution of the reconstructed mass of the candidates from the
 four experiments, compared with the expectation from the background
 and a potential signal with $\mh=115~\Gcs$, is shown in figure
 \ref{fig:massplots}(a).

 The signal-to-background ratio computed from the simulated samples,
 in the reconstructed mass region $m_{\mathrm{rec}}>109\,\Gcs$, is
 approximately 0.3.

 It is worth pointing out again the oft-repeated shortcomings of such
 a mass plot: the calculation of the likelihood ratio is based on more
 than just this distribution. Also, in such a plot --where all events
 have equal weight independently of e.g., which search channel or
 centre-of-mass energy they originate from-- it is impossible to
 convey the fact that different channels, centre-of-mass energies, can
 have widely different sensitivities to the signal hypothesis.

 In order to partially overcome such criticism and glimpse which are
 the candidates that have more impact on the calculation, one can
 tighten the event selection without biasing the mass
 distribution. This is achieved by, for instance, tightening the
 selection cut on the b-flavour probability of the events. Figure
 \ref{fig:massplots}(b) shows the high purity subset of candidates for
 which the signal-to-background ratio, in the reconstructed mass
 region $m_{\mathrm{rec}}>109\,\Gcs$, is 2.0. In the high mass region
 one can see an excess of candidates with respect to the SM
 background.
%
%
The four most significant candidates, in terms of their weight, are
three four-jet candidates from ALEPH\,\cite{aleph00} and one L3
candidate in the missing energy channel\,\cite{l3-nov2000}. Table
\ref{tab:candlist} lists the most significant candidates ordered 
by their contribution to the log-likelihood ratio\,\cite{run-request}.

\begin{table}
\caption{
List of the most significant candidates from the four LEP experiments
ordered by $s/b$ at $\mh=115\,\Gcs$.\label{tab:candlist}}
\vspace{0.2cm}
\begin{center}
\begin{tabular}{|ccccc|}
\hline
 Channel & Experiment & $\sqrt{s}$ & Reconstructed & $(\frac{s}{b})_{115}$\\ 
 & & (GeV) & mass ($\Gcs$) & \\
\hline
 $\h\qq$ & A & 206.7 & 114 & 4.7 \\
 $\h\qq$ & A & 206.7 & 112 & 2.3 \\
 $\h\nn$ & L & 206.6 & 114 & 2.05 \\
 $\h\qq$ & A & 206.7 & 110 & 0.90 \\
 $\h\ee$ & A & 205.3 & 118 & 0.60 \\
 $\h\qq$ & O & 205.4 & 113 & 0.52\\
 $\h\tt$ & A & 208.1 & 115 & 0.5 \\
 $\h\qq$ & A & 206.5 & 114 & 0.5 \\
 $\h\nn$ & L & 208.2 & 114 & 0.49\\
 $\h\qq$ & L & 206.7 & 115 & 0.47\\
 $\h\qq$ & D & 206.7 & 97  & 0.45\\
 $\h\qq$ & D & 206.7 & 114 & 0.40 \\
\hline
\end{tabular}
\end{center}
\end{table}

\begin{figure}[htb]
\begin{center}
\epsfig{file=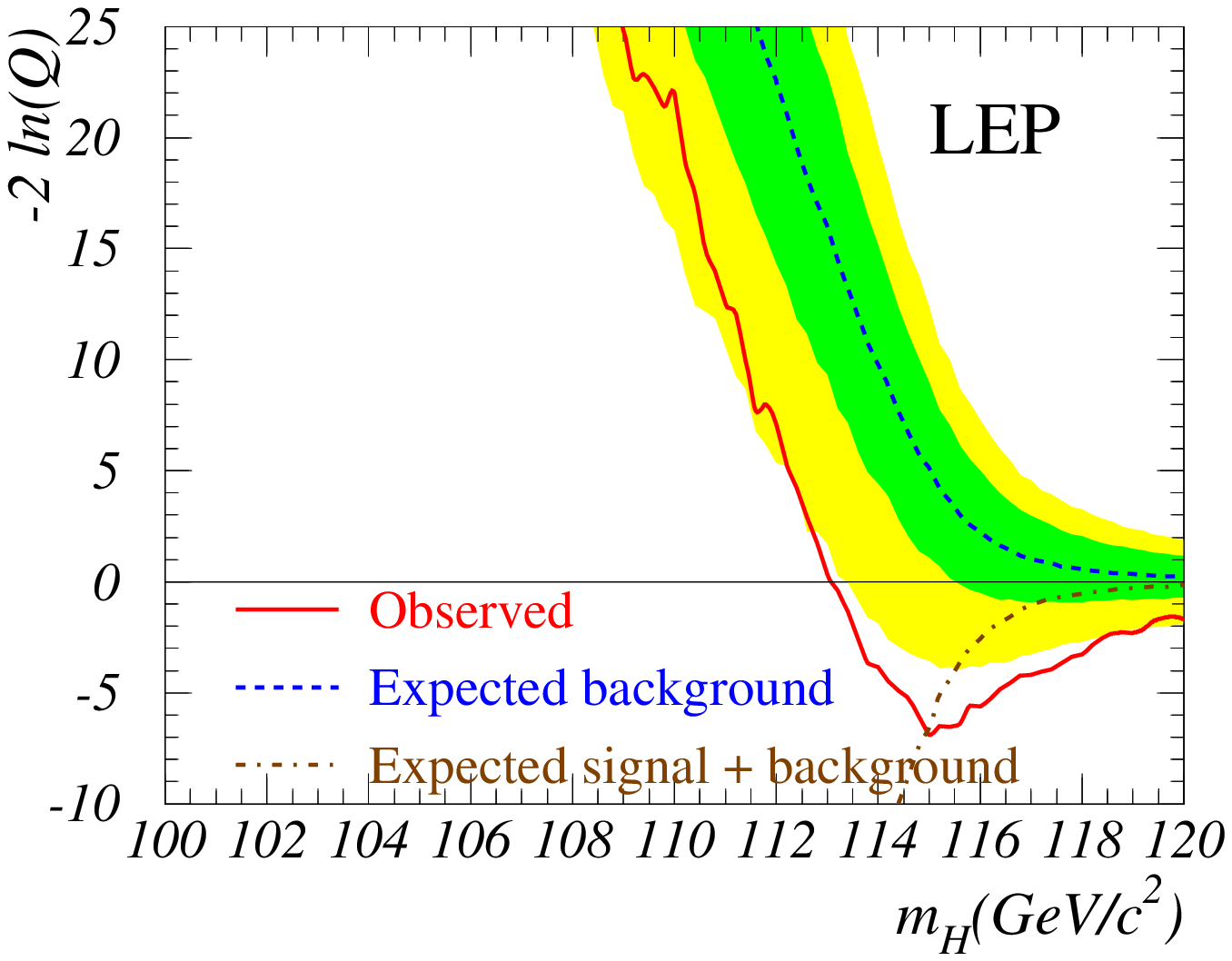,width=7.9cm} \epsfig{file=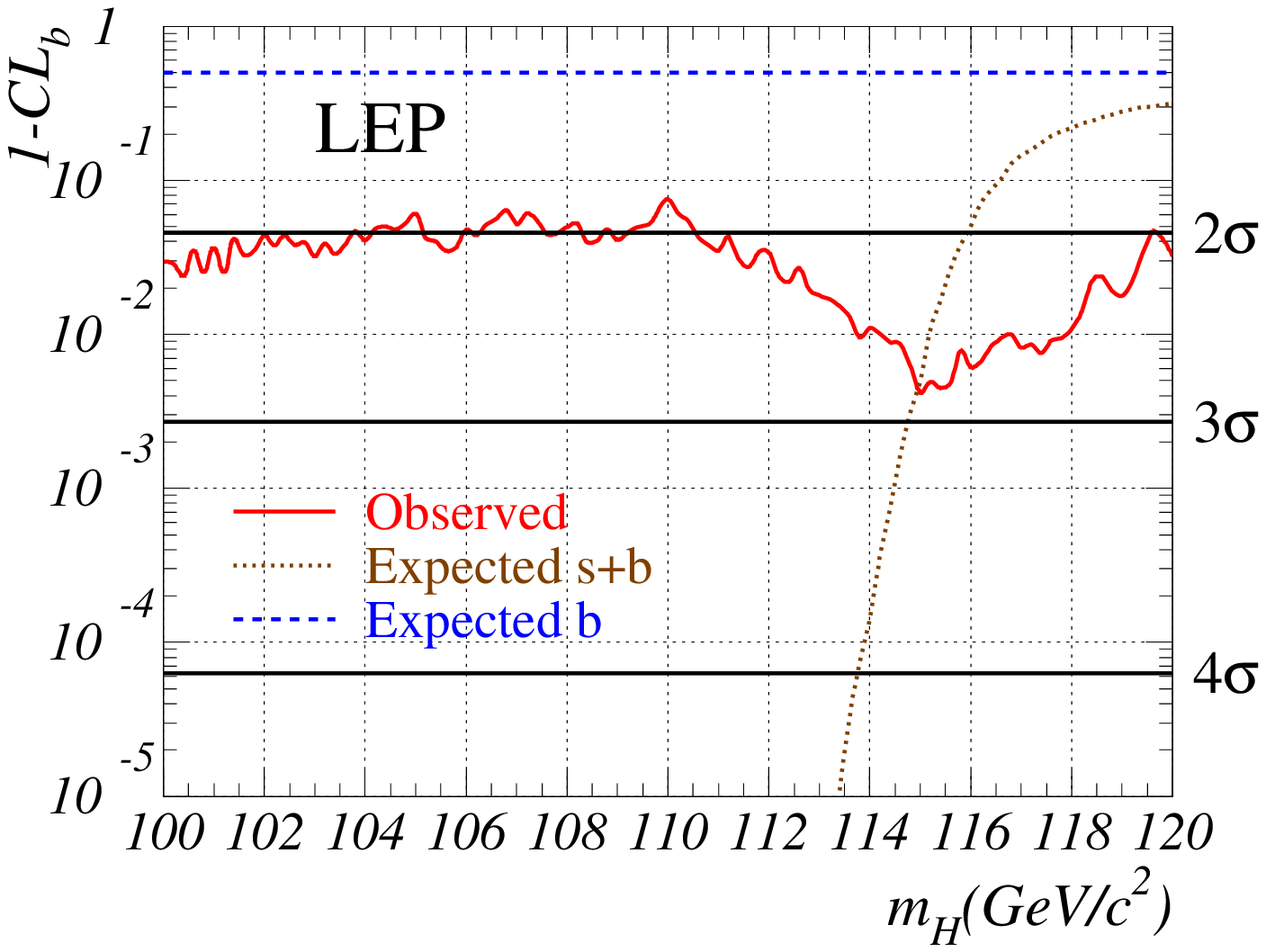,width=7.9cm} \\
\end{center}
\caption{The combined results from LEP. 
(a) The log-likelihood ratio curve as a function of the test mass
$\mh$. The solid line is the result obtained from the data. The dashed
line is the expected median in the background-only scenario. The
light- and dark-gray bands contain 68\% and 95\% of the simulated
background-only experiments. The dot-dashed line is the expected
position of the median log-likelihood when the latter is calculated at
a mass $\mh$ and includes a signal at that same mass.  (b) The
probability $1-CL_b$ that a fluctuation of the background produces a
result at least as signal-like as observed as a function $\mh$, for
the data (solid line) and the expected background (dashed line). $CL_b$
is the confidence level in the background hypothesis. The dotted line
indicates the location of the median for a Higgs signal of mass $\mh$.
\label{fig:results}}
\end{figure}

The log-likelihood ratio curve is shown in Figure
\ref{fig:results}(a). It can be clearly seen that the data favour the
signal+background hypothesis over the background only hypothesis, at
$\mh=115\,\Gcs$.

The probability at any given test mass $\mh$, that a fluctuation of
the background produces an event configuration at least as signal-like
as the one observed is shown in Figure \ref{fig:results}(b). For
$\mh=115\,\Gcs$ this probability is $4.2\times 10^{-3}$, corresponding
to a $2.9\sigma$ excess over the background.

\section*{{\underline{Note added (July 2001)}}}
At the time of submitting these proceedings, a new preliminary LEP
combined result has been released by the LEP Higgs working
group\,\cite{ptd-jamboree}. The current combined result corresponds to
an excess over the background at the $2\sigma$ level. The maximum
consistency with an eventual signal occurs at $\mh=115.6\,\Gcs$. In
summary, the main differences in the inputs with respect to the
combined result presented in these proceedings are:

\begin{itemize}
\item the L3 collaboration has released new search results\,\cite{l3-latest}
that supersede the earlier publication\,\cite{l3-nov2000}. The overall
L3 search sensitivity has improved by $1\,\Gcs$, mostly due to
improvements to the four-jet search. The missing energy search has
been revised. As a consequence the highest weight candidate from L3,
in the missing energy channel, saw its weight reduced by a factor
$\sim 2$. The overall significance of the L3 observation with respect
to the SM background processes was reduced from $\sim\,1.7\sigma$ to
$\sim\,1\sigma$.

\item the ALEPH collaboration updated its inputs to include a 2D correlation
correction in the four-jet channel. The overall significance of the
ALEPH excess over the background was reduced from $\sim\,3.4\sigma$ to
$\sim\,3.2\sigma$.

\item 
the data collected at $\sqrt{s}>$206\,GeV during the last days of LEP
running, and which had not been included in the November 2000
result\,\cite{PIKNov3}, has now been analysed by the four experiments
and included in the combination. The extra data totals
$\approx\,55$\,pb$^{-1}$.

\end{itemize}

The final combined result from LEP is expected towards the end of
2001, after the final publications of ALEPH, DELPHI and OPAL appear in
print.

\section*{Acknowledgements}
I would like to thank the conference organizers for inviting me to
attend a memorable physics meeting. The skiing was also great. Most
importantly, I thank all my colleagues in the LEP Higgs working group
which were involved in the preparation of the results presented here,
and which have assisted in the preparation of my talk.

\end{document}